# Nonextensibility of internal energy in incomplete statistics and the zeroth law of thermodynamics

Zhifu Huang and Jincan Chen*

Department of Physics and Institute of Theoretical Physics and Astrophysics, Xiamen University, Xiamen 361005, People's Republic of China

**Abstract.** On the basis of the entropy of incomplete statistics (IS) and the joint probability factorization condition, two controversial problems existing in IS are investigated, where one is what the correct expression of the internal energy for a composite system is and the other is whether the zeroth law of thermodynamics is true or not. Some new equivalent expressions of the internal energy of a composite system are derived through a precise mathematical calculation. Moreover, a self-consistent calculation is used to expound that the zeroth law of thermodynamics is also suitable for IS, but it can't be proven from theory. Finally, it is pointed out that the generalized zeroth law of thermodynamics and the expressions of the internal energy of a composite system derived in literature are incorrect, because two irrational assumptions have been implicitly introduced.

---

*Email address: jcchen@xmu.edu.cn

***Introduction.*** Incomplete statistics (IS) proposed by Wang [1] and Tsallis' statistics [2, 3] have been two important branches of nonextensive statistical mechanics [4-17]. Recently, IS has been used to research the thermostatistic properties of a variety of physical systems with long-range interacting and/or long-duration memory and many significant results have been obtained [6-16]. For example, it has been found that for some chaotic systems evolving in fractal phase space [7, 8], the entropy change in time due to the fractal geometry is assimilated to the information growth through the scale refinement; and that the generalized fermion distributions based on incomplete information hypothesis can be useful for describing correlated electron systems [9,10]. However, when investigating the thermostatistic properties of a composite system, specially for nonextensive systems with different q indices [8, 15], one has to deal with two fundamental problems which is similar to those in Tsallis' statistics solved recently [18], where one is how to give correctly the expression of internal energy of a composite system and the other is how to expound whether the zeroth law of thermodynamics is true or not. Although the two important problems have been discussed for many years [1, 6, 8, 11], they have not been solved up to now, and consequently, have affected the development and improvement of IS. Recently, a concomitant definition of the physical temperature in IS is given [16], which may lay a foundation for deeply investigating the problems existing in IS. In the present paper, with the help of the results obtained in Refs.[1, 16] and the joint probability factorization condition, we will discuss the two important problems mentioned above and give some useful conclusions.

***The temperature in IS.*** According to the results of IS proposed by Wang [1], the entropy for a nonextensive system may be expressed as

$$S_q = k \frac{\sum_{i=1}^{w} p_i - 1}{q-1} \qquad (1)$$

with the incomplete normalization

$$\sum_{i}^{w} p_i^q = 1, \qquad (2)$$

where $k$ is the Boltzmann constant, $p_i$ is the probability of the state $i$ among $W$ possible ones that are accessible to the calculation, $\varepsilon_i$ is the energy of the system at state $i$, q is the nonextensive parameter. For the sake of convenience, $\sum_{i=1}^{w}$ is replaced by $\sum_{i}$ below.

Using Eqs. (1) and (2) and the expression of internal energy

$$U_q = \sum_{i}^{w} p_i^q \varepsilon_i, \qquad (3)$$

one can obtain the distribution function [1, 4, 6, 11, 13, 15]

$$p_i = \frac{[1-(1-q)\beta\varepsilon_i]^{1/(1-q)}}{Z_q} \qquad (4)$$

and the expression of entropy [1, 4, 6, 11]

$$S_q = k \frac{Z_q^{q-1} - 1}{q-1} + k\beta Z_q^{q-1} U_q \qquad (5)$$

with

$$Z_q = \left\{ \sum_{i} [1-(1-q)\beta\varepsilon_i]^{\frac{q}{1-q}} \right\}^{1/q}. \qquad (6)$$

On the basis of the above results, it has been strictly proven that [16]

$$\sum_{i} p_i = Z_q^{q-1}[1-(1-q)\beta U_q] \qquad (7)$$

and

$$\partial S_q / \partial U_q = k\beta Z_q^{q-1}/q = k\beta' = \frac{1}{T}, \tag{8}$$

where $T$ is the physical temperature of the system in equilibrium. This $T$ definition is different from original one of IS in Refs. [1, 4, 6, 8, 11, 13, 15] where $\beta$ is defined as the physical (measurable) temperature through the zeroth law. It shows that the Lagrange multiplier $\beta$ in Refs. [1, 4, 6, 8, 11, 13, 15] is not equal to $1/(kT)$.

***Nonextensive expressions of the internal energy.*** For an independent system $C$ composed of two subsystems $A$ and $B$ of which the distributions satisfy [1-4, 6, 8, 11, 15, 18]

$$p_{ij}(C) = p_i(A)p_j(B) \tag{9}$$

or

$$p_{ij}^q(C) = p_i^q(A)p_j^q(B), \tag{10}$$

one can derive the pseudo-additivity entropy rule [1, 4, 6, 8, 11, 15]

$$S_q(C) = S_q(A) + S_q(B) + [(q-1)/k]S_q(A)S_q(B) \tag{11}$$

from Eqs. (1) and (9). Using the law of entropy conservation $\delta S_q(C) = 0$ and Eq. (11), one can obtain [6, 8, 10, 11, 15]

$$[1+(q-1)S_q(B)/k]\frac{\partial S_q(A)}{\partial U_q(A)}\delta U_q(A) + [1+(q-1)S_q(A)/k]\frac{\partial S_q(B)}{\partial U_q(B)}\delta U_q(B) = 0. \tag{12}$$

On the other hand, from Eqs. (4) and (7)-(9), we can obtain

$$\frac{[1-(1-q)\beta(C)\varepsilon_{ij}(C)]^{1/(1-q)}}{Z_q(C)} = \frac{[1-(1-q)\beta(A)\varepsilon_i(A)]^{1/(1-q)}[1-(1-q)\beta(B)\varepsilon_j(B)]^{1/(1-q)}}{Z_q(A)Z_q(B)}, \tag{13}$$

$$\frac{1-(1-q)\beta(C)U_q(C)}{Z_q^{1-q}(C)} = \frac{[1-(1-q)\beta(A)U_q(A)][1-(1-q)\beta(B)U_q(B)]}{Z_q^{1-q}(A)Z_q^{1-q}(B)}, \tag{14}$$

or

$$\frac{\beta'(C)}{\beta(C)} - (1-q)\beta'(C)U_q(C) = q[\frac{\beta'(A)}{\beta(A)} - (1-q)\beta'(A)U_q(A)][\frac{\beta'(B)}{\beta(B)} - (1-q)\beta'(B)U_q(B)]. \qquad (15)$$

Equations (13)-(15) are equivalent each other. It is seen from Eqs. (13)-(15) that the internal energy in IS is nonextensive.

It is significant to note the fact that in the derivative process of Eqs. (13)-(15), we do not add any assumption except Eqs. (1)-(3) and (9) which have been adopted in IS, and that Eqs. (13)-(15) are completely different from the nonextensive expressions of the internal energy derived in literature, because some irrational assumptions had been added in the calculative process (see discussion) in Refs.[1, 4, 6, 8, 11] .

*The zeroth law of thermodynamics.* When the systems A and B are in equilibrium, one important condition

$$T(A) = T(B) = T(C) \qquad (16)$$

may be adopted, and consequently, Eq. (15) may be simplified as

$$\frac{1}{\beta(C)} - (1-q)U_q(C) = q\beta'[\frac{1}{\beta(A)\beta(B)} - \frac{1-q}{\beta(A)}U_q(B) - \frac{1-q}{\beta(B)}U_q(A) + (1-q)^2 U_q(A)U_q(B)]. \qquad (17)$$

Using the law of energy conservation $\delta U_q(C) = 0$ and Eq. (17), one can obtain

$$[(1-q)U_q(B) - \frac{1}{\beta(B)}]\frac{\partial}{\partial U_q(A)}[\frac{\beta'}{\beta(A)(q-1)} + \beta' U_q(A)]\delta U_q(A)$$
$$+ [(1-q)U_q(A) - \frac{1}{\beta(A)}]\frac{\partial}{\partial U_q(B)}[\frac{\beta'}{\beta(B)(q-1)} + \beta' U_q(B)]\delta U_q(B) = 0 \qquad (18)$$

Substituting Eqs. (5), (7), (8) and (16) into Eq. (18), one has

$$\sum_j p_j(B)\delta U_q(A) + \sum_i p_i(A)\delta U_q(B) = 0. \qquad (19)$$

From Eqs. (1), (12) and (19), one obtains

$$\frac{\partial S_q(A)}{\partial U_q(A)} = \frac{\partial S_q(B)}{\partial U_q(B)} \quad \text{or} \quad \beta'(A) = \beta'(B), \qquad (20)$$

which is just the zeroth law of thermodynamics. Obviously, the physical essence of Eq. (20) is completely identical with that of Eq. (16). It implies the fact that starting from Eq. (16), one gets Eq. (20), which is the same result as Eq. (16). It is thus clear that the derivative process of Eq. (20) is only of a self-consistent calculation, but is not a proof for the zeroth law of thermodynamics in IS. Like in Tsallis' statistics [18], the zeroth law of thermodynamics still holds in IS, but it cannot be proved from theory. The conclusion conforms to Abe's standpoint [17], i.e., statistical mechanics may be modified but thermodynamics should remain unchanged.

*Discussion.* When one assumption [4]

$$Z_q(C) = Z_q(A)Z_q(B) \tag{21}$$

is adopted, Eqs. (13) and (14) may be, respectively, simplified as

$$\beta(C)\varepsilon_{ij}(C) = \beta(A)\varepsilon_i(A) + \beta(B)\varepsilon_j(B) + (q-1)\beta(A)\beta(B)\varepsilon_i(A)\varepsilon_j(B) \tag{22}$$

and

$$\beta(C)U_q(C) = \beta(A)U_q(A) + \beta(B)U_q(B) + (q-1)\beta(A)\beta(B)U_q(A)U_q(B). \tag{23}$$

From Eqs. (7), (8) and (21), one can obtain

$$\sum_{ij} p_{ij}(C) + (1-q)q\beta'(C)U_q(C)$$
$$= [\sum_i p_i(A) + (1-q)q\beta'(A)U_q(A)][\sum_j p_j(B) + (1-q)q\beta'(B)U_q(B)]. \tag{24}$$

Substituting Eqs. (9) and (16) into Eq.(24) gives

$$U_q(C) = \sum_i p_i(A)U_q(B) + \sum_j p_j(B)U_q(A) + (1-q)q\beta'U_q(A)U_q(B). \tag{25}$$

Using Eqs. (7) and. (25) and the law of energy conservation, we obtain

$$\left[\frac{\partial Z_q^{q-1}(A)}{\partial U_q(A)}U_q(B) + \sum_j p_j(B)\right]\delta U_q(A) + \left[\frac{\partial Z_q^{q-1}(B)}{\partial U_q(B)}U_q(A) + \sum_i p_i(A)\right]\delta U_q(B) = 0 \tag{26}$$

where [16]

$$\frac{\partial Z_q}{\partial U_q} = \frac{Z_q}{q} \frac{\sum_i e_q^{2q-1}(-\beta \varepsilon_i) \varepsilon_i}{\sum_i e_q^{2q-1}(-\beta \varepsilon_i)(\varepsilon_i^2 - U_q \varepsilon_i)} \tag{27}$$

and $e_q(x) = [1+(1-q)x]^{1/(1-q)}$. Equation (26) is obviously in contradiction with Eq. (12) because $\partial Z_q / \partial U_q \neq 0$, so that Eq. (20) can't be derived. This means that the calculative process is not self-consistent and that the assumption given above, i.e., Eq. (21), is not true.

If the other assumption

$$\beta(C) = \beta(A) = \beta(B), \tag{28}$$

is also adopted, Eqs. (22) and (23) may be further simplified as

$$\varepsilon_{ij}(C) = \varepsilon_i(A) + \varepsilon_j(B) + (q-1)\beta \varepsilon_i(A) \varepsilon_j(B) \tag{29}$$

and

$$U_q(C) = U_q(A) + U_q(B) + (q-1)\beta U_q(A) U_q(B), \tag{30}$$

respectively. Equations (29) and (30) are just some main results obtained in Refs. [1, 4, 6, 8, 11] and have been used to discuss the zeroth law of thermodynamics.

In Refs. [4, 11], Eq. (30) was used to calculate the variation of the internal energy

$$\partial U_q(C) = [1+(q-1)\beta U_q(B)]\partial U_q(A) + [1+(q-1)\beta U_q(A)]\partial U_q(B) \tag{31a}$$

$$\partial U_q(C) = [1+(q-1) U_q(B)/k]\partial U_q(A) + [1+(q-1) U_q(A)/k]\partial U_q(B) \tag{31b}$$

and to derive the generalized zeroth law of thermodynamics [4, 6, 11]

$$Z_q^{1-q}(A) \frac{\partial S(A)}{\partial U(A)} = Z_q^{1-q}(B) \frac{\partial S(B)}{\partial U(B)} \quad \text{or} \quad \beta(A) = \beta(B). \tag{32}$$

It seems to be a self-consistent calculation. However, it can be clearly seen to analyze Eqs. (30) and (31) that Eq. (31) is error so that Eq. (32) can't be derived. The cause may be explained as follows: when the systems are in equilibrium, the rational assumption should be described by Eq. (16) rather than Eq. (28), because $\partial S_q / \partial U_q = 1/T \neq k\beta$ [16].

If Eq. (30) were directly used, one would obtain the expression of the variation of the internal energy as

$$\delta U_q(C) = \{1+(q-1)[\beta U_q(B)+U_q(A)U_q(B)\partial\beta/\partial U_q(A)]\}\delta U_q(A)$$
$$+\{1+(q-1)[\beta U_q(A)+U_q(A)U_q(B)\partial\beta/\partial U_q(B)]\}\delta U_q(B) \qquad (33)$$

with [16]

$$\frac{\partial U_q}{\partial \beta} = \frac{q\sum_i e_q^{2q-1}(-\beta\varepsilon_i)(U_q\varepsilon_i - \varepsilon_i^2)}{Z_q^q}. \qquad (34)$$

Using the law of entropy conservation $\delta U_q(C) = 0$ and Eqs. (12) and (33), one only obtains the following equation

$$\sum_j p_j(B)\{1+(q-1)[\beta U_q(A)+U_q(A)U_q(B)\partial\beta/\partial U_q(B)]\}\frac{\partial S_q(A)}{\partial U_q(A)}$$
$$= \sum_i p_i(A)\{1+(q-1)[\beta U_q(B)+U_q(A)U_q(B)\partial\beta/\partial U_q(A)]\}\frac{\partial S_q(B)}{\partial U_q(B)}. \qquad (35)$$

Equation (35) is obviously different from Eq. (32) and is in contradiction with Eq. (28). It indicates that it is incorrect to introduce the generalized zeroth law of thermodynamics.

The above discussion shows clearly that two additional assumptions described above are irrational and have been implicitly added to the calculative process in Refs. [1, 4, 6, 8, 11], so Eqs. (29) and (30) are not the corrective expressions of the internal energy in IS.

*Conclusions*. With the help of the entropy expression proposed by Wang [1] and the results obtained in Ref. [16], we have solved two important problems in IS. The internal energy in IS is nonextensive. Its corrective expressions should be given by Eqs. (13)-(15) rather than Eqs. (29) and (30). The zeroth law of thermodynamics can't be proved from theory, but it is still true in IS, while the so-called generalized zeroth law of thermodynamics may not be correct. The results obtained here shows clearly that two irrational additional assumptions have been implicitly used in some relevant literatures.

**Acknowledgments:**

This work has been supported by the Research Foundation of Ministry of Education, People's Republic of China.